\documentclass[12pt]{article}
\usepackage{amsmath}

\renewcommand{\d}{{\rm d}}
\newcommand{\bl}{\mbox{\boldmath$l$}}
\newcommand{\bN}{\mbox{\boldmath$N$}}
\newcommand{\bv}{\mbox{\boldmath$v$}}
\newcommand{\btau}{\mbox{\boldmath$\tau$}}
\newcommand{\pR}{\,^{+}\!R}

\newcommand{\pOmega}{\,^{+}\!\Omega}
\newcommand{\mOmega}{\,^{-}\!\Omega}
\newcommand{\gsimeqscr}{\begin{smallmatrix} > \\ \sim \end{smallmatrix}}
\newcommand{\lsimeqscr}{\begin{smallmatrix} < \\ \sim \end{smallmatrix}}
\newcommand{\vf}{\mbox{\boldmath$f$}}
\newcommand{\bx}{\mbox{\boldmath$x$}}
\newcommand{\by}{\mbox{\boldmath$y$}}
\newcommand{\bp}{\mbox{\boldmath$p$}}
\newcommand{\bDelta}{\overline{\Delta}}
\newcommand{\bn}{\mbox{\boldmath$n$}}
\newcommand{\bJ}{\mbox{\boldmath$J$}}
\newcommand{\tilS}{\tilde{S}}

\begin{document}

\title{On the discrete version of the Schwarzschild problem
}

\author{V.M. Khatsymovsky \\
 {\em Budker Institute of Nuclear Physics} \\ {\em of Siberian Branch Russian Academy of Sciences} \\ {\em
 Novosibirsk,
 630090,
 Russia}
\\ {\em E-mail address: khatsym@gmail.com}}
\date{}
\maketitle
\begin{abstract}
We consider a Schwarzschild type solution in the discrete Regge calculus formulation of general relativity quantized within the path integral approach.
Earlier, we found a mechanism of a loose fixation of the background scale of Regge lengths. This elementary length scale is defined by the Planck scale and some free parameter of such a quantum extension of the theory.
Besides, Regge action was reduced to an expansion over metric variations between the tetrahedra and, in the main approximation, is a finite-difference form of the Hilbert-Einstein action.
Using for the Schwarzschild problem a priori general non-spherically symmetrical ansatz, we get finite-difference equations for its discrete version.
This defines a solution which at large distances is close to the continuum Schwarzschild geometry, and the metric and effective curvature at the center are cut off at the elementary length scale.
Slow rotation can also be taken into account (Lense-Thirring-like metric).
Thus we get a general approach to the classical background in the quantum framework in zero order: it is an optimal starting point for the perturbative expansion of the theory; finite-difference equations are classical, the elementary length scale has quantum origin. Singularities, if any, are resolved.
\end{abstract}


MSC classes: 83C27; 83C57

keywords: Einstein theory of gravity; minisuperspace theory; piecewise flat space-time; Regge calculus; Schwarzschild black hole

\section{Introduction}

The task of studying the object indicated by the title of the article is a special case of an attempt to describe a system with extreme and even singular gravitational fields. Such a description, as is generally accepted, requires the involvement of quantum gravity and, in turn, can be considered in the broader context of studying quantum effects in the framework of a specific quantum-gravitational approach.

From the formal viewpoint, general relativity (GR) is a non-renormalizable field theory, and divergences originate from the continuum nature of space-time. This leads us to expect the efficiency of discrete approaches \cite{Ham}. A distinctive feature of gravity as a geometry consists in the presence of a simple ansatz of geometry, which is described by a discrete set of variables. This is a piecewise flat manifold which can be viewed as a collection of flat 4-dimensional tetrahedra or 4-simplices. Regge calculus is the GR on the class of such piecewise flat manifolds \cite{Regge}. These manifolds can approximate any given Riemannian geometry with arbitrarily high accuracy \cite{Fein,CMS}. This makes it possible to use such a formulation in quantum applications, usually using a functional integral approach as well. A certain freedom is connected with the choice of a functional measure, which can be made based on reasonable physical assumptions and used to extract physical quantities such as the Newtonian potential \cite{HamWil1,HamWil2}.

A (geometry of the) piecewise flat manifold can be fully characterized by the edge lengths. These edge lengths can be viewed as the result of a {\it triangulation} of some manifold. Regge calculus corresponds to an intuitive "experimental" viewpoint on the active role of the measurement process creating a state: we can only measure a discrete (more exactly, arbitrarily large finite) set of triangulation data, and we only have, in a sense, the simplest geometry corresponding to this data.

In a more general context, there are various discrete approaches to gravity based on triangulation. Spin foam models \cite{Per} generalize to four dimensions the work by Ponzano-Regge in three dimensions \cite{Pon}, where the edge lengths are considered to be quantized as moments, and state sum over triangulations is that of the products of the 6j-symbols for the tetrahedra and is shown to correspond to an effective action relating approximately to the (three-dimensional) Regge action. Here the state sum is primary, and the action is secondary. In the more traditional for the field theory approach we have taken, the (Regge) action is primary, and the path integral is constructed from it. There are also options here. In the Causal Dynamical Triangulations theory (CDT) \cite{cdt}, several different building blocks (4-simplices) are considered. We stick to the conventional Regge calculus ideology, which assumes continuous changes in the edge lengths. In the paper \cite{Mik}, the effective action formalism was applied to the Regge path integral to obtain the possibility to have the observed cosmological constant value.

An application of the Regge calculus to static charged and uncharged black holes was considered classically in the paper \cite{Wong}. A piecewise flat manifold breaks spherical symmetry, and if a classical fixed simplicial structure is used, it is important to minimize this breaking. Regge calculus was applied to the three-dimensional space, which was replaced by a certain icosahedral three-dimensional simplicial structure. Besides that, for the numerical study of such systems, effective lattice methods are proposed that are alternative to the Regge calculus \cite{Bre2}. In a broader context, Regge calculus was used to numerically analyze cosmological models \cite{WilCol} -- \cite{WilLiu}. In the Causal Dynamical Triangulations theory, the emergence of cosmological models was considered \cite{GlaLol}.

As concerning studying a quantum black hole, the Loop Quantum Gravity has been applied to the Schwarzschild problem \cite{Ash1,Ash2}. Although this is not a discrete theory, its connection with discrete gravity models was considered \cite{Dup}, and, besides, the singularity is resolved just due to the discreteness of the area/length spectrum (and, therefore, the finiteness of the area/length quantum in this theory).

In the approach we have taken, with the edge lengths as independent variables, it is important to have a mechanism for dynamically loose fixing these lengths around a certain scale \cite{our1}. This mechanism arises due to a specific form of measure in the functional integral formalism. The most direct route to the functional integral is through the canonical Hamiltonian formalism and quantization. However, if such a formalism is formulated in terms of only edge lengths, it is singular. A way out of the situation appears when using the representation of the Regge action in terms of edge lengths and connection SO(3,1) matrices as independent variables. Thus, we get a certain analogue of the situation in the continuum GR, where we can develop the canonical formalism, proceeding from the first order Cartan-Weyl connection form of the action. Having obtained the functional integral in terms of the edge lengths and the SO(3,1) connection matrices, we can integrate over the latter and end up with the functional integral only in terms of the edge lengths. The resulting functional integral measure in terms of edge lengths has the ability to loosely fix an elementary length scale. This manifests itself in the form of a fixation of an optimal starting point (in the configuration superspace) of the perturbative expansion series for this functional integral, like the fixation of the equilibrium point in a potential well.

As a result, we have conditions on the initial point of the perturbative expansion in the form of a condition for some maximization of the measure (this determines the elementary length scale) and, as usual, the equation of motion or the Regge equation. In \cite{our2}, we calculate the Regge action on a simplicial complex, on which some coordinates of the vertices are given, with the help of the intermediate use of {\it discrete} Christoffel symbols or GL(4.R) matrices relating the affine frames of the neighboring 4-simplices. In this way, the action can be represented as a series in typical metric variations between neighboring 4-simplices, the leading term for the simplest periodic simplicial structure being a finite-difference form of the Hilbert-Einstein action. Thus, in the leading order over metric variations, we come to the analysis of the finite-difference Einstein equations.

Since a finite-difference form breaks the spherical symmetry (as well as any other continuum symmetry), passing to the finite differences should be made without a priori substituting a spherically symmetrical ansatz for the metric into the equations. Rather this should be the general ten-component metric. Though, the general form of the solution to the Einstein equations is not known yet. But it turns out that one can restrict oneself in the leading order over metric variations from 4-simplex to 4-simplex by the three-component vector part of these ten degrees of freedom. Then the equations are solvable, including their finite-difference form. The present paper continues our paper \cite{Kha1} where a non-rotating black hole was considered. Now we generalize this to the case of a slow rotating black hole (the discrete version of the Lense-Thirring problem), the methodology is given in more detail, some points are worked out.

The paper is organized as follows. The method is considered in Section \ref{method}. The mechanism for a dynamical loose fixation of the edge lengths implies using discrete lapse-shift functions being constant parameters; a particular choice of the appropriate space-time simplicial structure is analogous to that one following by triangulating the continuum space-time in the synchronous frame coordinates. In Section \ref{calculation}, the calculation is described. In the leading order over metric variations from 4-simplex to 4-simplex used, we can pass to the formulas for the case of a simplicial structure analogous to that following by triangulating the continuum space-time in some other coordinate system. The latter is chosen in Subsection \ref{equations} to correspond to the aforementioned three-component vector ansatz, and the discrete equations and metric solution are considered. In Subsection \ref{Lense-Thirring}, the discrete equations and metric solution are considered in the case of a slow rotation of the body. Then Discussion follows.

\section{The method}\label{method}

Eventually, the problem is reduced to solving the classical Regge skeleton equations, but the elementary length scale in our approach has a quantum origin, and we explain how this problem arises formally in the functional integral framework. Namely, this boils down to the problem of finding the optimal starting point (that is, specific configuration (s)) of the formal perturbative expansion for the functional integral in terms of the edge lengths (solving the equations of motion (\ref{dS/dl=0}) and (\ref{f^2/(d^2s/dldl)=max}) for the maximum of the functional measure below).

Since extending the configuration superspace by including connection SO(3,1) matrices as independent variables helps in the non-singular canonical quantization of (the continuous time form of) the Regge calculus, in particular, in the path integral form, constructing the path integral in terms of both the edge lengths (or, more accurately, edge vectors) and connection matrices was taken as the initial one. The functional integral in terms of the edge lengths follows as a result of integration over connection. (In principle, such an integration can be explicitly performed term by term for the expansion over the discrete lapse-shifts \cite{Kha2}.) This is equivalent to a partial summation of the perturbative diagrams (those with the internal lines being connection ones). The resulting measure possesses the maximum providing the loose fixation of the edge lengths at certain values, or, roughly, at a certain scale $a$ (\ref{a=sqrt}) \cite{our1}. This $a$ is proportional to the Planck length and thus, in the usual units, to $\sqrt{\hbar }$. It is interesting that in the classical limit $\hbar \to 0$ we have $a \to 0$, that is, passing to the continuum. In other words, discreteness here is a quantum effect.

In somewhat more detail, we consider a piecewise flat space-time represented by a simplicial complex consisting of 4-dimensional tetrahedra or 4-simplices $\sigma^4$, usual tetrahedrons as their 3-dimensional faces $\sigma^3$, triangles as the 2-dimensional faces $\sigma^2$, edges $\sigma^1$ and vertices $\sigma^0$. Given such a simplicial structure, the geometry is fully characterized by the set of the edge lengths $\ell \equiv (l_1, \dots, l_n )$ (depending on the context, $\ell $ can denote a set of edge vectors, see below). Regge action is written as a sum over triangles $\sigma^2$ (2-simplices) in terms of their areas $A_{\sigma^2}$ and {\it angle defects} on them $\alpha_{\sigma^2}$ (the difference between $2 \pi$ and the sum of the hyper-dihedral angles meeting at $\sigma^2$), functions of $\ell$,
\begin{equation}\label{Regge}                                               
S ( \ell ) = \frac{1}{8 \pi G} \sum_{\sigma^2} A_{\sigma^2} \left ( \ell \right ) \alpha_{\sigma^2} \left ( \ell \right )
\end{equation}

\noindent (we use the units, in which $\hbar = 1 $, $c = 1$).

For the possibility of non-singular canonical quantization, as mentioned, it is useful to extend the set of independent variables by including some variables of the connection type $\Omega$ such that excluding these from a certain action form $S(\ell , \Omega  )$ via equations of motion would result in $S( \ell )$ (\ref{Regge}),
\begin{equation}\label{S(l,Omega)=S(l)}                                     
\partial_{\Omega} S(\ell , \Omega  ) = 0 \Rightarrow S(\ell , \Omega ( \ell ) ) = S ( \ell ) .
\end{equation}

\noindent To this end, a local pseudo-Euclidean frame is assigned to each 4-simplex. Each edge $\sigma^1$ is characterized by a vector $l^a_{\sigma^1}$ in the frame of a certain $\sigma^4 \supset \sigma^1$ (a more detailed notation being $l^a_{\sigma^1 | \sigma^4}$). The notions of discrete tetrad and connection were introduced by Fr\"{o}hlich \cite{Fro}. It is also important to be able to restrict oneself to the (anti-)self-dual parts of the connection matrices for better computability and in order to be able to write (a discrete version of) the parity odd Holst term \cite{Holst,Fat}, parameterized by the Barbero-Immirzi parameter $\gamma$ \cite{Barb,Imm}, in order, in particular, to correspond to the discrete version of the system initial for obtaining the Loop Quantum Gravity formulation. We have suggested in \cite{Kha} such a form $S(\ell , \Omega  )$ in terms of area tensors and finite rotation SO(4) (SO(3,1) in the considered Minkowsky case) matrices, and also in terms of (anti-)self-dual parts of finite rotation matrices. The connection matrices $\Omega_{\sigma^3}$ "live" on the tetrahedra $\sigma^3$ (3-simplices). The so-called "continuous time limit" can be performed, and this can be recast in the canonical Hamiltonian form with some Lagrangian $L = \sum_{\sigma^2} [l_1, l_2 ] \overline{\Omega} \dot{\Omega } - H(\ell , \Omega  )$ (symbolically), $[l_1, l_2 ]$ being a bivector of a triangle $\sigma^2$. The canonical quantization can be performed in the functional integral form $\int \exp [ i S(\ell , \Omega  ) ] \d \mu ( \ell ) {\cal D} \Omega$. (The functional integral measure can be chosen so that it reduces to the canonical quantization measure in the continuous time limit whatever coordinate is taken as a time.) After integrating over $\Omega$, we are left with a functional integral only over the length variables, $\int \exp [ i \tilS ( \ell ) ] F ( \ell ) D \ell$, with a phase $\tilS ( \ell )$ and a module $F ( \ell )$,
\begin{equation}                                                            
\int \exp [ i S(\ell , \Omega  ) ] \Psi ( \ell ) \d \mu ( \ell ) {\cal D} \Omega = \int \exp [ i \tilS ( \ell ) ] \Psi ( \ell ) F ( \ell ) D \ell
\end{equation}

\noindent with a probe function $\Psi ( \ell )$.

For $\tilS ( \ell )$, it is appropriate to use the stationary phase expansion, for a nonzero phase appears already in the zero order. In this order, it follows by excluding $\Omega$ classically from $S(\ell , \Omega  )$ and is just the Regge action $S( \ell )$ by definition of $S(\ell , \Omega  )$ (\ref{S(l,Omega)=S(l)}).

For $F ( \ell )$, it is appropriate to use an expansion over the discrete analogs of the Arnowitt- Deser-Misner (ADM) \cite{ADM1} lapse-shift functions $(N, N^i)$, for a nontrivial module appears just in the zero order of this expansion. More exactly, such discrete analogs can be introduced for a certain type of the simplicial structure. Namely, it is assumed to be constructed from analogous neighboring in time three-dimensional simplicial complexes (the leaves or slices of the foliation) using {\it temporal} and {\it diagonal} edges. A temporal edge connects a pair of analogous vertices in two neighboring leaves. A diagonal edge connects a vertex $\sigma^0_1$ in a leaf with a vertex $\sigma^0_{2^\prime}$ in a neighboring leaf whose analogue $\sigma^0_2$ in the former leaf is a neighbor of $\sigma^0_1$ ($\sigma^0_1$, $\sigma^0_2$ are the ends of some edge $(\sigma^0_1 \sigma^0_2 )$). The vector $l^a_{\sigma^1}$ of a temporal edge $\sigma^0_1$ is just a discrete analogue of $N^a = (N, \bN)$, and we can denote $l^a_{\sigma^1} \equiv N^a_{\sigma^1} = (N_{\sigma^1}, \bN_{\sigma^1})$. In such a structure, we can distinguish between a {\it spatial} triangle (completely contained in a leaf), a {\it temporal} triangle (having a temporal edge) and a {\it diagonal} triangle which is neither of these two. (We previously referred to "temporal edge" and "spatial edge" as to "$t$-like edge" and "leaf edge" respectively \cite{our1}.) The edge vector $l^a_{\sigma^1}$ is an analogue of $e^a_\lambda$ of the continuum theory, and the terms "temporal $\sigma^1$" and "spatial $\sigma^1$" are analogs of the "covariant world vector index $\lambda = 0$" and "$\lambda = 1, 2, 3$", respectively. Whereas "time-like" and "space-like" refer to the "local vector index $a = 0$" and "$a = 1, 2, 3$", respectively. Typical spatial, temporal and diagonal edges are shown in Fig.~\ref{triangulation} below. This classification is important when expanding the functional integral over the discrete lapse-shift functions,
\begin{eqnarray}\label{int-d-Omega-over-N}                                  
& & \int \exp \left \{ \frac{i}{2} \left \{ \left ( 1 + \frac{i}{\gamma } \right ) \left [ \sum_{\stackrel{{\scriptstyle\rm spatial/dia-}}{{\rm gonal~}\sigma^2}} \sqrt{ \bv^2_{\sigma^2} } \arcsin \frac{\bv_{\sigma^2} * \pR_{\sigma^2} ( \Omega )}{\sqrt{ \bv^2_{\sigma^2}}} \right. \right. \right. \nonumber \\ & & \left. \left. \left. + \sum_{{\rm temporal~}\sigma^2} \sqrt{ \btau^2_{\sigma^2} } \arcsin \frac{\btau_{\sigma^2} * \pR_{\sigma^2} ( \Omega )}{\sqrt{ \btau^2_{\sigma^2}}} \right ] + \underset{\textstyle \rm conjugate }{\rm complex } \right \} \right \} {\cal D} \Omega \nonumber \\ & & \hspace{-10mm} = \int \exp \left \{ \frac{i}{2} \left [ \left ( 1 + \frac{i}{\gamma } \right ) \sum_{\stackrel{{\scriptstyle\rm spatial/dia-}}{{\rm gonal~}\sigma^2}} \sqrt{ \bv^2_{\sigma^2} } \arcsin \frac{\bv_{\sigma^2} * \pR_{\sigma^2} ( \Omega )}{\sqrt{ \bv^2_{\sigma^2}}} + \underset{\textstyle \rm conjugate }{\rm complex } \right ] \right \} {\cal D} \Omega \nonumber \\ & & + O(N) .
\end{eqnarray}

\noindent The connection representation used is based on the decomposition of SO(3,1) as (a subgroup of) SO(3,C) $\times$ SO(3,C), self-dual $\times$ anti-self-dual ones, $\Omega = \pOmega \mOmega , \mOmega = ( \pOmega )^* , \pR ( \Omega ) = R ( \pOmega )$.
The curvature $R_{\sigma^2} ( \Omega ) = \prod_{{\sigma^3} \supset {\sigma^2}} \Omega^{\pm 1}_{\sigma^3}$, holonomy of $\Omega$; $\pR$ is represented here as a 3 $\times$ 3 matrix. The area bivector of the triangle $\sigma^2 = [\sigma^1_1 \sigma^1_2]$ (formed by two edges $\sigma^1_1 $, $\sigma^1_2$) is characterized by the complex area 3-vector $\bv_{\sigma^2}$, $2 \bv_{\sigma^2} = i \bl_{\sigma^1_1} \times \bl_{\sigma^1_2} - \bl_{\sigma^1_1} l^0_{\sigma^1_2} + \bl_{\sigma^1_2} l^0_{\sigma^1_1}$; for the temporal triangle ($\sigma^1_1$ being a temporal edge), this vector is denoted as $\btau_{\sigma^2}$: $2 \btau_{\sigma^2} = i \bN_{\sigma^1_1} \times \bl_{\sigma^1_2} - \bN_{\sigma^1_1} l^0_{\sigma^1_2} + \bl_{\sigma^1_2} N_{\sigma^1_1}$. For a 3-vector and a 3 $\times$ 3 matrix, $\bv * R \equiv \frac{1}{2}v^i R^{kl} \epsilon_{ikl}$. The quantity $\gamma$ is the Barbero-Immirzi parameter. In (\ref{int-d-Omega-over-N}), the contribution of the temporal triangles is singled out. It is of order $N^a$, and in zero order in $N^a$, the functional integral over $\Omega$ is calculable ($R_{\sigma^2}$ included in the expression can be taken as independent variables).

The measure $F ( \ell )$ is defined up to $V^\eta$ ($V$ is the 4-simplex volume), an analogue of $(\sqrt{-g})^\eta$ in the continuum GR. (In the continuum GR, different choices of $\eta$ lead, eg, to the DeWitt measure \cite{DeW} or to the Misner measure \cite{Mis}.) $F ( \ell )$ turns out to have a maximum at the triangle areas $\propto a^2 / 2$,
\begin{equation}\label{a=sqrt}                                              
a = \sqrt{ 32 G ( \eta - 5) / 3 } .
\end{equation}

\noindent ($\sqrt{G}$ is the Planck length).

The aforementioned expansions for the action $\tilS ( \ell )$ and for the measure $F ( \ell )$ can both be regarded as expansions over small parameters if $\ell$ is loosely fixed at the scale $a \gg 1$ in the Planck units (this means $\eta \gg 1$); also then we can take the Regge action $S (\ell )$ as $\tilS (\ell )$.

More strictly, the fixation of $\ell$ is finding an optimal starting point $\ell = \ell_0$ of the perturbative expansion. Let $F ( \ell ) D \ell = D u$ be the Lebesgue measure in some new $u = (u_1, \dots, u_n )$,
\begin{eqnarray}\label{int-exp-iS-dl}                                       
& & \int \exp [ i S ( \ell ) ] F ( \ell ) D \ell = \int D u \exp \left \{ i \left [ S ( \ell_0 ) \right. \right. \nonumber \\ & & \hspace{-10mm} \left. \left. + \frac{1}{2} \sum_{jklm} \frac{\partial^2 S (\ell_0 )}{\partial l_j \partial l_l} \frac{\partial l_j (u_0 )}{\partial u_k} \frac{\partial l_l (u_0 )}{\partial u_m} (u - u_0)_k (u - u_0)_m + \dots \right ] \right \} .
\end{eqnarray}
To define the point $\ell_0 = \ell (u_0 )$, it is insufficient to use the equations of motion,
\begin{equation}\label{dS/dl=0}                                             
\frac{\partial S(\ell_0)}{\partial \ell} = 0
\end{equation}
(finding an extremum of the zero order term), but also there is the possibility to require an extremum of the second order term,
\begin{equation}\label{f^2/(d^2s/dldl)=max}                                 
F (\ell_0 )^2 \det \left \| \frac{\partial^2 S (\ell_0 )}{\partial l_i \partial l_k} \right \|^{-1} = \mbox{maximum}.
\end{equation}
The point $\ell = \ell_0$ is defined by (\ref{dS/dl=0}), (\ref{f^2/(d^2s/dldl)=max}). Note that the system of (\ref{dS/dl=0}) and (\ref{f^2/(d^2s/dldl)=max}) is invariant in its form if we pass to new variables $\ell^\prime$ related to $\ell$ in a nondegenerate way $\ell \to \ell^\prime = \ell^\prime ( \ell )$.

The matrix $\partial^2 S (\ell_0 ) / \partial l_i \partial l_k$ has zero order in the scale of edge lengths. This matrix is close to a diagonal one (only those $l_i$ and $l_k$ "interact" in the Regge action $S$ which refer to the same 4-simplex); at the same time, geometrically, the edge length scale can not change abruptly from simplex to simplex. Therefore, it is expected that the inclusion of the determinant of this matrix in (\ref{f^2/(d^2s/dldl)=max}) will not lead to an essential change in the extreme point $\ell_0$ of (\ref{f^2/(d^2s/dldl)=max}) compared to the maximum of only $F (\ell_0 )$. This also means some sufficient uniformity of the elementary length scale.

How to solve $S(l_0)/\partial l = 0$?  We can expand $S(l_0)$ over the metric variations between the 4-simplices \cite{our2}. In the simplest periodic simplicial structure with a 4-cube cell divided by diagonals into 4!=24 4-simplices \cite{RocWil}, we introduce coordinates which run through the fours of integers $(n_1, n_2, n_3, n_4)$ at the vertices and for which the metric is constant inside each 4-simplex.

The leading term turns out to be a finite-difference form of the Hilbert-Einstein action in terms of the metric variations between the {\it 4-cubes},
\begin{eqnarray}\label{DM+MM}                                                
\sum_{\rm 4-cubes} {\cal K}^{\lambda \mu}_{~~~ \lambda \mu} \sqrt{g} , ~~~ {\cal K}^\lambda_{~ \, \mu \nu \rho} \! = \! \Delta_\nu M^\lambda_{\rho \mu} \! - \! \Delta_\rho M^\lambda_{\nu \mu} \! + \! M^\lambda_{\nu \sigma} M^\sigma_{\rho \mu} \! - \! M^\lambda_{\rho \sigma} M^\sigma_{\nu \mu} , \nonumber \\ \hspace{-10mm}
M^\lambda_{\mu \nu} = \frac{1}{2} g^{\lambda \rho} (\Delta_\nu g_{\mu \rho} + \Delta_\mu g_{\rho \nu} - \Delta_\rho g_{\mu \nu}), ~~~ \Delta_\lambda = 1 - \overline{T}_\lambda .
\end{eqnarray}
$T_\lambda$ is the shift operator along the coordinate $x^\lambda$ by 1.

We aim to consider the piecewise flat manifold of interest as a starting point of the perturbative expansion.

The (finite-difference) field equations are classical, the elementary length scale $a$, at which the measure has the maximum, has a quantum nature.

When calculating the functional measure, an expansion over the discrete lapse-shift functions as parameters is used; these can be taken by hand as analogs of the continuum gauge parameters. A particular case is the synchronous frame $N = 1$, $N^i = 0$. Or, here, a discrete metric close to the Lemaitre one \cite{Lemaitre}.

\section{Calculation}\label{calculation}

In the Lemaitre type coordinates $r_1$ \cite{Stan}, $\tau$, $r^{3/2} = r^{3/2}_1 - \frac{3}{2} \sqrt{r_g} \tau $,
\begin{equation}\label{Lemaitre}                                           
\hspace{-0mm} \d s^2 \! = \! - \! \d \tau^2 \! + \! \frac{r_1}{r(r_1, \tau )} \! \d r^2_1 \! + \! r^2 (r_1 , \tau ) \d \Omega^2 \! = \! - \! \d \tau^2 \! + \! ( \d r(r_1, \tau ) |_{\tau = const} )^2 \! + \! r^2 (r_1 , \tau ) \! \d \Omega^2.
\end{equation}

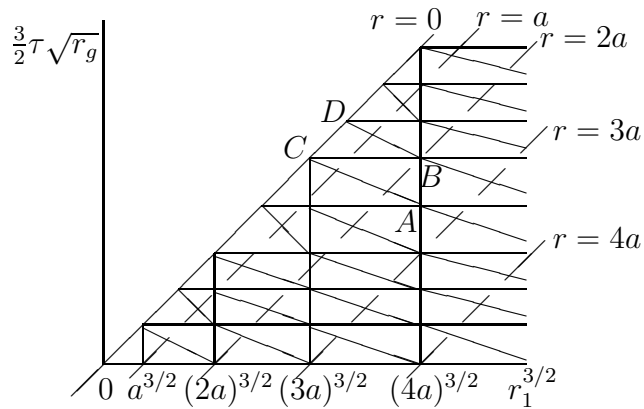
\begin{figure}[h]
\unitlength 1pt
\begin{picture}(120,152)(-140,-62)
\put(-43,-43){\line(-1,-1){12}}
\put(-43,-43){\line(1,0){160}}
\put(-43,-43){\line(0,1){130}}
\put(-43,-43){\line(1,1){125}}
\put(110,-56){$r_1^{3/2}$}
\put(65,-56){$(4a)^{3/2}$}
\put(23,-56){$(3a)^{3/2}$}
\put(-13,-56){$(2a)^{3/2}$}
\put(-34,-56){$a^{3/2}$}
\put(-45,-56){$0$}
\put(-78,75){$\frac{3}{2} \tau \sqrt{r_g}$}
\put(-28,-43){\line(0,1){15}}
\put(-28,-29){\line(2,-1){27}}
\put(-14.5,-14.5){\line(1,-1){13.5}}
\put(-1,-43){\line(0,1){42}}
\put(-1,-28){\line(5,-2){36}}
\put(-1,-15.5){\line(3,-1){36}}
\put(-1,-2){\line(3,-1){36}}
\put(17,17){\line(1,-1){18}}
\put(35,-43){\line(0,1){78}}
\put(35,-29){\line(3,-1){42}}
\put(35,-14.5){\line(3,-1){42}}
\put(35,-1){\line(3,-1){42}}
\put(35,16){\line(5,-2){42}}
\put(35,34.5){\line(5,-2){42}}
\put(49,49){\line(2,-1){28}}
\put(25,35){$C$}
\put(39,49){$D$}
\put(67,7){$A$}
\put(76,24){$B$}
\put(77,-43){\line(0,1){120}}
\put(77,-28){\line(3,-1){40}}
\put(77,-14.5){\line(4,-1){40}}
\put(77,-1){\line(4,-1){40}}
\put(77,17){\line(3,-1){40}}
\put(77,35){\line(3,-1){40}}
\put(77,49){\line(4,-1){40}}
\put(77,63){\line(4,-1){40}}
\put(77,77){\line(4,-1){40}}
\put(-28,-28){\line(1,0){145}}
\put(-1,-1){\line(1,0){118}}
\put(-14.5,-14.5){\line(1,0){131.5}}
\put(35,35){\line(1,0){82}}
\put(17,17){\line(1,0){100}}
\put(77,77){\line(1,0){40}}
\put(63,63){\line(1,0){54}}
\put(49,49){\line(1,0){68}}
\put(63,63){\line(1,-1){14}}
\put(-33,-48){\line(1,1){15}}
\put(-12,-27){\line(1,1){10}}
\put(4,-11){\line(1,1){10}}
\put(20,5){\line(1,1){10}}
\put(36,21){\line(1,1){10}}
\put(52,37){\line(1,1){10}}
\put(68,53){\line(1,1){10}}
\put(84,69){\line(1,1){15}}
\put(-6,-48){\line(1,1){15}}
\put(15,-27){\line(1,1){10}}
\put(31,-11){\line(1,1){10}}
\put(47,5){\line(1,1){10}}
\put(63,21){\line(1,1){10}}
\put(79,37){\line(1,1){10}}
\put(95,53){\line(1,1){10}}
\put(111,69){\line(1,1){10}}
\put(30,-48){\line(1,1){15}}
\put(51,-27){\line(1,1){10}}
\put(67,-11){\line(1,1){10}}
\put(83,5){\line(1,1){10}}
\put(99,21){\line(1,1){10}}
\put(115,37){\line(1,1){10}}
\put(58,84){$r=0$}
\put(98,85){$r=a$}
\put(122,77){$r=2a$}
\put(127,42){$r=3a$}
\put(72,-48){\line(1,1){15}}
\put(93,-27){\line(1,1){10}}
\put(109,-11){\line(1,1){15}}
\put(127,1){$r=4a$}
\end{picture}
\caption{A typical triangulation with the simplest periodic lattice in the Lemaitre coordinates, drawn in a section passing through the world line $r = 0$ (however, for another such section, there are events of ending geodesics $r_1 = const$ at $r = 0$, say, $r_1 = a \sqrt{2} $, which do not belong to any of these leaves $\tau = const$; therefore, the structure should be distorted near $r=0$, for example, a line $r_1 = const$ before $r = 0$ can end with a non-($r_1 = const$) edge). $BC$, $AB$ and $AC$ are examples of the spatial, temporal, and diagonal edges, respectively. $C D$ is an edge at $r=0$.}\label{triangulation}
\end{figure}

As an example, we can consider a simplicial structure with 4-cube cells, whose spatial bases are in the (flat) 3D leaves $\tau = const$, temporal edges correspond to geodesic lines $r_1 = const$, orthogonal to these leaves, and their time-like length is $\Delta \tau$ (the difference between the neighboring leaves $\tau = const$), Fig.~\ref{triangulation}, where the edge lengths are the lengths of the corresponding geodesic segments.

A temporal edge vector, including its length $\Delta \tau$, is fixed by hand analogously to the continuum lapse-shift functions, which are gauge functions. Triangulation can be referred to as a measurement procedure (including quantum one), fixing $\Delta \tau$ as a definition of this procedure. Once the elementary length scale is dynamically fixed at $a$, the choice $\Delta \tau \ll a$ means an overestimation versus the quantum uncertainty $a$, $\Delta \tau \gg a$ an insufficient detail compared to the quantum uncertainty $a$, and the essential deviation of $\Delta \tau / a$ from unity in both these cases makes the description of the system somewhat singular. An important part of the Regge calculus strategy is averaging over various simplicial structures. As a part of such an averaging, averaging over different temporal edge lengths seems to be appropriate, but now we can take for estimate $\Delta \tau \simeq a$, which also reflects a symmetry between space and time.

Note that in some applications it is preferable that all edge lengths are not null \cite{Wil}; writing $\Delta \tau \simeq a$ leaves room for this if accidentally (say, at $\Delta \tau = a$, $\Delta r = a$, $r_g = 0$) the length of the edge turns out to be this null. Although here we do not explicitly use this opportunity; only the existence of triangulation is important.

We can estimate the region in which the metric variations are small, and, thus, the skeleton equations can be accurately approximated by the finite-difference ones. Since local relations (differential equations in the finite-difference form) are studied, we can limit ourselves to an interval from $\tau = 0$ to some $\tau \sim a $ in order to have a few 3D leaves sufficient to form a 4-geometry. In the region of interest, deviation of the metric between $\tau = 0$ and $\tau \simeq a$ from the flat one, which is defined by $1 - r / r_1$, should be small,
\begin{equation}\label{Dr<<r1}                                             
\left | \frac{r_1 (r, \tau ) - r}{r_1 (r, \tau )} \right | \ll 1 \mbox{ at } r_g \ll \frac{r^3 }{a^2 } .
\end{equation}

\noindent Alternatively, we can issue from a typical value of the angle defect $\alpha \sim R a^2$ on the elementary area scale $a^2$ for the typical curvature in the continuum GR $R \sim r_g / r^3$ (from the curvature invariants). The condition for the smallness of the angle defect gives the same estimate,
\begin{equation}                                                           
\alpha \sim a^2 R \sim a^2 \frac{r_g}{r^3} \ll 1 \mbox{ at } r_g \ll \frac{r^3 }{a^2 } .
\end{equation}

In particular, the above estimate can be prolonged to $r \gsimeqscr a$ (the nearest to $r=0$ vertex) at
\begin{equation}\label{r_g<<a}                                             
r_g \ll a ,
\end{equation}

\noindent though this is not physically quite an interesting case (there is no horizon as such). Note that $a \gg 1$ at $\eta \gg 1$ (see (\ref{a=sqrt})), and $r_g \gg 1$ admits (\ref{r_g<<a}).

The metric in the 4-simplices/cubes substituted in the finite-difference form (\ref{DM+MM}) can be viewed as the values of some smooth (interpolating) field $g_{\lambda \mu}$ in the 4-simplices or cubes, and in the above region (\ref{Dr<<r1}), the Lemaitre metric can be taken for $g_{\lambda \mu}$.

In the leading order over the finite differences, the latter obey the rules for the derivatives, and the finite-difference expression for the action possesses the invariance with respect to redefining the coordinates of the vertices. So we can go to another coordinate system in this order.

\subsection{Eguations in the leading order over metric variations}\label{equations}

The most convenient seems to be a discrete analogue of a certain generalization of the Painlev\'{e}-Gullstrand metric \cite{Painleve,Gullstrand}, for it allows us to formulate the Schwarzschild problem without the a priori requirement of spherical symmetry (to which no lattice obeys).

The Painlev\'{e}-Gullstrand metric
\begin{equation}\label{Pan}                                                
\d s^2 = - \d \tau^2 + \left ( \d r + \sqrt{\frac{r_g}{r}} \d \tau \right )^2 + r^2 \d \Omega^2 .
\end{equation}

\noindent This follows at $f^k = x^k \sqrt{r_g /r^3} $ from
\begin{equation}                                                           
\d s^2 = - \d \tau^2 + \sum^3_{k = 1} (\d x^k + f^k \d \tau)^2 ,
\end{equation}

\noindent which does not have spherical symmetry for the general $f^k$. This fits naturally into the general 3+1 ADM form of metric
\begin{equation}\label{3+1}                                                
\d s^2 = - ( N \d \tau )^2 + g_{k l} (\d x^k + f^k \d \tau ) (\d x^l + f^l \d \tau ) .
\end{equation}

\noindent In the discrete case, $N - 1$ and $g_{k l} - \delta_{k l}$ are not zero, but the next-to-leading order in the metric variations $O(\delta )$.

We can generalize the procedure of passing from the Painlev\'{e}-Gullstrand (\ref{Pan}) to the Lemaitre (\ref{Lemaitre}) metric to the case of the general 3+1 ADM form of metric (\ref{3+1}). Finding this change of variables $( \bx , \tau ) \to ( \by , \tau )$, $x^k = x^k ( \by , \tau )$ amounts to solving the differential equations
\begin{eqnarray}\label{dx/dt+f=0}                                       
& & \frac{\partial x^k ( \by , \tau )}{\partial \tau } + f^k ( \bx (\by , \tau ) , \tau ) = 0 , ~~~ x^k (\by , 0 ) = y^k \\ & & \mbox{ or } ~~~ x^k ( \by , \tau ) = y^k - \int^\tau_0 f^k ( \bx (\by , \tau ) , \tau ) \d \tau
\end{eqnarray}

\noindent in the integral form. Then the metric reads
\begin{equation}\label{likeLemaitre}                                       
\d s^2 = - ( N \d \tau )^2 + g_{k l} \frac{\partial x^k }{\partial y^m } \frac{\partial x^l }{\partial y^n } \d y^m \d y^n .
\end{equation}

\noindent In principle, according to our strategy of having fixed constant discrete lapse-shifts (required to perform the functional integral expansion over them), we could try to perform a transformation to achieve $N = 1$. However, the continuum version already has $N = 1$, and it is quite unreal that its discrete analogue would have abnormally large discrete lapse-shifts; the bounded on the whole space-time discrete lapse-shifts will do as well. After all, in the end we turn to another coordinates and only the fact of the existence of such a metric is important.

The finite-difference version of the GR action (\ref{DM+MM}) can be calculated on the metric ansatz (\ref{likeLemaitre}). Approximate relations for finite differences are useful here, for example,
\begin{equation}\label{chain}                                              
\frac{\Delta g_{kl} (\bx (\by , \tau ) , \tau )}{\Delta y^m} = \frac{\Delta x^n ( \by , \tau )}{\Delta y^m} \frac{\Delta g_{kl} (\bx (\by , \tau ) , \tau )}{\Delta x^n (\by , \tau )} ,
\end{equation}

\noindent for the discrete version of the chain rule for the derivative of a composite function, of the product rule and so on. These relations are the more accurate, the smaller are the variations of the metric from simplex to simplex. Indeed, the estimate of the discrepancy when approximating the typical derivative by finite differences appearing when passing from $r_1$ (a function of $\by$) to $r$ (a function of $\bx$) gives that it is small at
\begin{equation}\label{Dr1-dr1<<1}                                         
\left | \frac{a}{r_1 (r + a, \tau ) - r_1(r, \tau )} - \frac{\partial r}{\partial r_1(r + a, \tau )} \right | \left | \frac{\partial r}{\partial r_1(r + a, \tau )} \right |^{-1} \ll 1 \mbox{ at } r_g \ll \frac{r^3 }{a^2 } ,
\end{equation}

\noindent that is, when (\ref{Dr<<r1}) holds. (Note that the minimal nonzero $r$ of a vertex in a leaf is $a$.) A similar requirement for the relative accuracy of the approximation of $\partial^2 r / \partial r_1^2$ by $\Delta^2 r / \Delta r_1^2$ leads to the condition $r >> a$. This second derivative vanishes at $r_g = 0$, which corresponds to the fact that the action is nonzero due to the part of the metric that vanishes at $r_g = 0$ ( $O(\sqrt{r_g} )$ ), and, therefore, the accuracy should be estimated relative to this part. The condition $r >> a$ is in fact that one that the derivatives of several negative powers of $r$, such as $r^{-1}$, can be approximated by the finite differences (and vice versa). In the physically interesting case $r_g > a$, the condition $r >> a$ is weaker than the condition $r \gg (a^2 r_g )^{1/3}$ (\ref{Dr<<r1}). In the case $r_g \lsimeqscr a$, the condition for the reliability of the leading order over metric variations is $r >> a$, and for $r \lsimeqscr a$, a semi-quantitative estimate can be expected.

In the leading order over metric variations, the finite differences in the form for the action can be handled as the corresponding derivatives. In particular, the general covariance holds, and the action in terms of the metric as a function of the coordinates $y^k$ can be rewritten in terms of the metric in the original coordinates $x^k$ upon substituting the metric ansatz (\ref{likeLemaitre}) into the action. Thus, a finite-difference form of the action in terms of the 3+1 ADM form of metric (\ref{3+1}) follows. Generally speaking, such a reduction takes place only in the leading order over metric variations. In the non-leading orders, the Taylor series corrections should be taken into account in the relation between finite differences and derivatives, and passing to another coordinates is not so simple. Moreover, the non-leading terms also come as corrections when reducing the original Regge action to the above finite-difference form (\ref{DM+MM}).

The action is \cite{ADM1} (from now on, we omit the factor $(16 \pi G )^{-1}$)
\begin{eqnarray}                                                           
& & S = \int \left \{ - g_{k l} \frac{\partial \pi^{k l}}{\partial \tau} + N \sqrt{g} \left [ ^3 R + g^{-1} \left ( \frac{1}{2} \pi^k{}_k \pi^l{}_l - \pi^k{}_l \pi^l{}_k \right ) \right ] + 2 f_k \pi^{k l}{}_{| l} \right \} \d^3 x \d \tau , \nonumber \\ & &  \pi_{k l} = \sqrt{g} (g_{k l} K^m{}_m - K_{k l}) , ~~~ K_{k l} = \frac{1}{2N} \left ( f_{k | l} + f_{l | k} - \frac{\partial g_{k l}}{\partial \tau} \right ) ,
\end{eqnarray}
$|$ in indices is the covariant differentiation with respect to $g_{k l}$.

Let us write down the field equations \cite{ADM1}. Those obtained by varying $S$ over $g_{kl}$,
\begin{equation}\label{dS/dg}                                              
0 = - \frac{1}{N \sqrt{g}} \frac{\delta S}{\delta g_{k l}} = ^3 \! R^{k l} - \frac{1}{2} g^{k l} \, {}^3 \! R + \dots ,
\end{equation}

\noindent can be written as
\begin{eqnarray}\label{3R-g3R/2}                                           
& & ^3 \! R^{k l} - \frac{1}{2} g^{k l} \, {}^3 \! R = - \frac{1}{2} \frac{g^{k l}}{g}  \left ( \frac{1}{2} \pi^m{}_m \pi^n{}_n - \pi^m{}_n \pi^n{}_m \right ) \nonumber \\ & & + \frac{2}{g} \left ( \frac{1}{2} \pi^m{}_m \pi^{k l} - \pi^k{}_m \pi^{m l} \right ) + \frac{1}{N \sqrt{g}} \left [ (\pi^{k l} f^m)_{| m} - f^k{}_{| m} \pi^{m l} - f^l{}_{| m} \pi^{m k} \right ]  \nonumber \\ & & + N^{-1} \left ( N^{| k l} - g^{k l} N^{| m}{}_{| m} \right ) - \frac{1}{N \sqrt{g}} \frac{\partial \pi^{k l}}{\partial \tau} .
\end{eqnarray}

\noindent They look like three-dimensional Einstein equations with a non-trivial dependence on $g_{kl}$ on the right-hand side. These six equations with respect to their left-hand side are not independent due to the Bianchi identities, only three components of them are independent; three more equations express the equality to zero of the divergence of the right-hand side. Formally, these six equations define $^3 \! R^{k l}$ in terms of the right-hand side and, therefore, $^3 \! R^{k l m n}$.

The equations obtained by varying $S$ over $f^k$,
\begin{equation}\label{dS/df}                                              
\hspace{-20mm} 0 = \frac{N}{\sqrt{g}} \frac{\delta S}{\delta f^k} = - (f_{k | l} - f_{l | k})^{| l} + \dots ,
\end{equation}

\noindent read
\begin{eqnarray}\label{ddf-ddf=}                                           
& & \hspace{-20mm} (f_{k | l} - f_{l | k})^{| l} = - 2 ^3 \! R_{k l} f^l + ( \ln N )^{| l} \left ( f_{k | l} + f_{l | k} -2 g_{kl} f^m_{| m} \right ) \nonumber \\ & & + \left ( g^{m n} \delta^l_k - g^{l m} \delta^n_k \right ) \left [ \left ( \frac{\partial g_{l m}}{\partial \tau} \right )_{| n} \! \! \! - (\ln N)_{| n} \frac{\partial g_{l m}}{\partial \tau} \right ] .
\end{eqnarray}

Varying $S$ over $N$, we see that the resulting equation includes $^3 \! R$. Finding this value from (\ref{3R-g3R/2}), as mentioned, we can exclude it from $\delta S / \delta N = 0$. Combining this also with $\delta S / \delta f^k = 0$ for a more compact dependence on $f_k$,
\begin{equation}\label{dS/dg-dS/df-dS/dN}                                  
0 = \! \frac{N}{\sqrt{g}} \left [ \frac{ g_{k l}}{2} \frac{\delta S}{\delta g_{k l}} - \! f^k \frac{\delta S}{\delta f^k} - \! \frac{ N}{4} \frac{\delta S}{\delta N} \right ] \! = \! \frac{1}{2} (f^k f_k)^{| l}_{| l} - \! \frac{1}{4} (f^{k | l} \! - \! f^{l | k} )(f_{k | l} \! - \! f_{l | k} ) + \dots ,
\end{equation}

\noindent we get
\begin{eqnarray}\label{dd(ff)=}                                            
& & \hspace{-5mm} \frac{1}{2} (f^k f_k)^{| l}_{| l} - \frac{1}{4} (f^{k | l} - f^{l | k} )(f_{k | l} - f_{l | k} ) = \nonumber \\ & & \hspace{-5mm} - ^3 \! R_{k l} f^k f^l + ( \ln N )_{| l} \left[ (f^k f_k)^{| l} - f_k (f^{k | l} - f^{l | k} ) - f^l f^k_{| k} \right] + N N^{| k}_{| k}
\nonumber \\ & & \hspace{-10mm} + \frac{1}{2} \frac{N}{\sqrt{g}} g_{k l} \frac{\partial \pi^{k l}}{\partial \tau} + \frac{1}{2} f^{k | l} \frac{\partial g_{k l}}{\partial \tau} + \left ( g^{m n} f^l - \frac{1}{2} g^{l m} f^n \right ) \left [ \left ( \frac{\partial g_{l m}}{\partial \tau} \right )_{| n} \! \! \! - (\ln N)_{| n} \frac{\partial g_{l m}}{\partial \tau} \right ] .
\end{eqnarray}

The resulting equations (\ref{3R-g3R/2}), (\ref{ddf-ddf=}), (\ref{dd(ff)=}) are in the form with the right-hand sides being zero on the Painlev\'{e}-Gullstrand metric. (This is seen from (\ref{ddf-ddf=}), (\ref{dd(ff)=}) at $^3 \! R_{k l} = 0$, $N = const$, $\partial g_{k l} / \partial \tau = 0$, $\partial f_k / \partial \tau = 0$ and checked by substituting the Painlev\'{e}-Gullstrand metric into (\ref{3R-g3R/2})).

The equations (\ref{ddf-ddf=}) are not independent with respect to their left-hand side (which is purely transversal); as a result, we have vanishing divergence of the right-hand side as a consistency condition,
\begin{eqnarray}\label{ddlnN=R}                                            
& & \left [ ( \ln N )^{| l} \left ( 2 g_{kl} f^m_{| m} - f_{k | l} - f_{l | k} \right ) \right ]^{| k} = \left \{ - 2 \, ^3 \! R_{k l} f^l +  \phantom{\left [ \left ( \frac{1}{2} \right )_{| k} \right ]} \right. \nonumber \\ & & \left. + \left ( g^{m n} \delta^l_k - g^{l m} \delta^n_k \right ) \left [ \left ( \frac{\partial g_{l m}}{\partial \tau} \right )_{| n} \! \! \! - (\ln N)_{| n} \frac{\partial g_{l m}}{\partial \tau} \right ] \right \}^{| k} .
\end{eqnarray}

\noindent This can be considered as an equation for $N$. This equation is non-degenerate with respect to its left-hand side at the point of the Painlev\'{e}-Gullstrand metric (as is seen by substitution of the corresponding $\vf$ into the left-hand side of (\ref{ddlnN=R})). Thus, equations (\ref{ddf-ddf=}) and (\ref{dd(ff)=}) can be considered as equations for $N$, $\vf$.

In overall, equations (\ref{3R-g3R/2}), (\ref{ddf-ddf=}), (\ref{dd(ff)=}), in which one equation from (\ref{ddf-ddf=}) should be substituted by the consistency condition (\ref{ddlnN=R}), and three equations from (\ref{3R-g3R/2}) should be substituted by certain three consistency conditions (for which we can take those that express the equality to zero of the divergence of the right-hand side), seem to be resolvable iteratively. With these caveats in mind, the equations have the form
\begin{eqnarray}                                                     
\label{6eqs} ^3 \! R^{k l} - \frac{1}{2} g^{k l} \, {}^3 \! R = O \left (N_{| l}, \frac{\partial \pi^{k l}}{\partial \tau} \right ) & & \mbox{(6 eqs)}, \\
\label{1eq} \nabla^2 \left (\vf^2 \right ) - \left [\nabla \times \vf \right ] \cdot \left [\nabla \times \vf \right ] = O \left (^3 \! R_{k l}, N_{| l}, \frac{\partial \pi^{k l}}{\partial \tau}, \frac{\partial g_{k l}}{\partial \tau} \right ) & & \mbox{(1 eq)}, \\
\left [\nabla \times [ \nabla \times \vf ] \right ] = O \left (^3 \! R_{k l}, N_{| l}, \frac{\partial g_{k l}}{\partial \tau} \right ) & & \mbox{(3 eqs)}.
\end{eqnarray}

In the continuum case of a spherically symmetrical (here without derivatives) $\delta$-function source, if we assume $N = 1$, $g_{k l} = \delta_{k l}$ and independence from $\tau$, or solve iteratively, starting with the Minkowski metric, then we get
\begin{eqnarray}                                                        
\label{1eq0} \nabla^2 \left (\vf^2 \right ) - \left [\nabla \times \vf \right ] \cdot \left [\nabla \times \vf \right ] = 0 & & \mbox{(1 eq)}, \\ \label{3dep-eqs} \left [\nabla \times [ \nabla \times \vf ] \right ] = 0 & & \mbox{(3 dependent eqs)}
\end{eqnarray}

\noindent at $r > 0$ and the Painlev\'{e}-Gullstrand metric.

It is interesting to express the individual equations (\ref{dS/dg}), (\ref{dS/df}), (\ref{dS/dg-dS/df-dS/dN}) in terms of the components of the Ricci/Einstein tensor. Whereas (\ref{dS/dg}) and (\ref{dS/df}) are $G^{k l}$ and $2 (G_{0 k} - G_{k l} f^l )$, respectively, (\ref{dS/dg-dS/df-dS/dN}) is more compact in terms of the covariant Ricci tensor components,
\begin{equation}                                                           
\frac{N}{\sqrt{g}} \left [ \frac{ g_{k l}}{2} \frac{\delta S}{\delta g_{k l}} - f^k \frac{\delta S}{\delta f^k} - \frac{ N}{4} \frac{\delta S}{\delta N} \right ] = - R_{0 0} + f^k f^l R_{k l} .
\end{equation}

\noindent The equations $\delta S / \delta N = 0$ and $\delta S / \delta f_k = 0$ are the equations for initial conditions in the Hamiltonian formalism, and $\delta S / \delta g_{k l} = 0$ are (a part of) the dynamical equations in this formalism. We see that a certain combination of the equations for initial conditions and the dynamical equations leads to some equations which can be called equations for $\vf , N$; the overall system with the $\delta$-function source and with $N = 1$, $g_{k l} = \delta_{k l}$ and independence from $\tau$ (for a similar definition of the Schwarzschild problem in the discrete case) or solved iteratively, starting with the Minkowsky metric (as an example for using, in principle, such a procedure to refine the solution in the discrete case), gives the Painlev\'{e}-Gullstrand metric.

In the discrete case in the leading order over $\delta g_{\lambda \mu}$, we have
\begin{equation}                                                           
f_k = a^{-1} \Delta_k \chi , ~~~ \sum^3_{k=1} \bDelta_k \Delta_k \left ( \vf^2 \right ) = \left \{ \begin{array}{rl} 0 & \mbox{at } \bx \neq 0  \\ C & \mbox{at } \bx = 0 , \end{array} \right.
\end{equation}

\noindent $\Delta_k h( x^k ) \equiv h(x^k ) - h(x^k - a )$, $C$ is chosen from requiring that $\vf^2$ be close to $r_g / r$ at $r \to \infty$ ($C = 4 \pi a^2 r_g$). We find
\begin{equation}\label{f^2}                                                
\vf^2 ( \bx ) = \int^{ \pi / a }_{\! \! - \pi / a } \int^{ \pi / a }_{\! \! - \pi / a } \int^{ \pi / a }_{\! \! - \pi / a } \frac{\d^3 \bp}{(2 \pi )^3} \frac{\pi a^2 r_g \exp (i \bp \, \bx )}{\sum_k \sin^2 (p_k a / 2 )}.
\end{equation}

\noindent Then the equation $a^{-2} \sum^3_{k=1} ( \Delta_k \chi )^2 = \vf^2$ should be solved and $f_k = a^{-1} \Delta_k \chi$ found. At the distances $r \gg a$, the contribution of the wave vectors up to $p \sim r^{-1}$ dominates, $\sum_k 4 \sin^2 (p_k a / 2 ) = a^2 p^2 + O(a^4 p^4 )$, the continuum function $\vf^2 = r_g r^{-1}$ is reproduced with a relative accuracy of the order of $(a / r)^2$. At $r = 0$, the cutoffs $| p_k | \leq \pi / a$ in (\ref{f^2}) are important, and $\vf^2 ({\bf 0} ) \sim r_g / a$ (more accurately, the values of (\ref{f^2}) at the center and at the nearest vertices are $\vf^2 ({\bf 0} ) \approx 1.05 \pi r_g / a$ and $\vf^2 (a, 0, 0 ) \approx 1.19 r_g / a$ \cite{Kha1} ). A convenient approximate form of the functions of interest can be given as follows,
\begin{equation}                                                           
f_k = a^{-1} \Delta_k \chi , ~~~ \chi ( \bx ) = 2 \sqrt{r_g r} \mbox{ at } r \geq a , ~~~ \chi ( {\bf 0} ) = 0.95 \sqrt{r_g a} .
\end{equation}

\noindent The corresponding $\vf^2$ reproduces $\vf^2$ (\ref{f^2}) at $\bx = {\bf 0}$ and approximates (\ref{f^2}) from above with an error less than 20\% at $r \geq a$ and decreasing to zero at $r \to \infty$. On the other hand, the continuum $\vf^2 = r_g r^{-1}$ approximates (\ref{f^2}) from below with an error less than 20\% at $r \geq a$ and diminishing to zero at $r \to \infty$; only at $\bx = {\bf 0}$ it cannot approximate, since it goes to infinity itself.

In the classical limit, $\hbar \to 0$, passing to the usual units, we get the continuum, $a \to 0$, as noted in Section \ref{method} (the end of the second paragraph). Accordingly, we get the expression for the continuum, $\vf^2 \to r_g r^{-1}$, and the value at the center depends on $\hbar$ nonanalytically, $\vf^2 ({\bf 0} ) \sim \hbar^{-1/2}$.

We also consider in our paper \cite{Kha1} an averaged version of metric or of any function of it. This is implied by the functional integral strategy, which presupposes averaging the result for any considered physical quantity over various simplicial structures. Limiting ourselves to the considered 4-cube lattice and the (defined by $\eta$) scale $a$, we can still average over the orientations of the lattice relative to the center and observation point $\bx$. This is a simplified model of this averaging. In particular, as applied to the metric function $\vf^2 ( \bx )$ itself, this reduces to averaging over the angle components of $\bx$ on the right-hand side of (\ref{f^2}) by applying $\int ( \cdot ) \d^2 \bn / (4 \pi )$, $\bx = r \bn$, $\bn^2 = 1$,
\begin{equation}\label{<f^2>}                                              
\overline{ \vf^2 ( \bx ) } = \int^{ \pi / a }_{\! \! - \pi / a } \int^{ \pi / a }_{\! \! - \pi / a } \int^{ \pi / a }_{\! \! - \pi / a } \frac{\d^3 \bp}{(2 \pi )^3} \frac{ \pi a^2 r_g}{\sum_k \sin^2 (p_k a / 2 )} \frac{\sin pr}{pr} .
\end{equation}

\noindent Of course, this is a function of $r$ only.

While we found that for $r \gg a$, the discrete solution found is formally close to the continuum one (that by Painlev\'{e}-Gullstrand), to be sure that the original skeleton equations are accurately approximated by the finite-difference ones, the condition $r \gg (a^2 r_g )^{1/3}$ (\ref{Dr<<r1}) should be fulfilled. In the case $r_g > a$, the latter is a stronger condition, and for $r \lsimeqscr (a^2 r_g )^{1/3}$, one should solve the original skeleton equations instead. In the formal case $r_g \lsimeqscr a$ (when in fact there is no horizon), as mentioned after (\ref{Dr1-dr1<<1}), the leading order over metric variations is reliable for $r >> a$, and a semi-quantitative estimate can be expected for $r \lsimeqscr a$.

We have estimated (\ref{f^2}) at $\bx = 0$ in our paper \cite{Kha1} (to be $r_g / a$ times a certain numerical constant). There we also estimated the discrete Riemannian tensor and, in particular, Kretschmann scalar $R_{\lambda \mu \nu \rho} R^{\lambda \mu \nu \rho}$ at the center in the formal case $r_g \ll a$ (to be $r_g^2 / a^6$ times a certain numerical constant).

\subsection{Discrete version of the Lense-Thirring metric}\label{Lense-Thirring}

We can also consider the case of a slowly rotating body at large distances. This system is described by the Lense-Thirring metric \cite{LenTir}, taking into account the rotation in the first (linear) approximation. This metric can be transformed to the Painlev\'{e}-Gullstrand type coordinates \cite{Visser},
\begin{eqnarray}\label{Lense}                                              
& & \d s^2 = - \d \tau^2 + \left ( \d r + \sqrt{\frac{r_g}{r}} \d \tau \right )^2 + r^2 \d \Omega^2 - \frac{4J}{r} \sin^2 \theta \d \varphi \d \tau \nonumber \\ & & \hspace{-10mm} = - \d \tau^2 + \left [  \d x + \left ( \frac{x}{r} \sqrt{\frac{r_g}{r}} + \frac{2J}{r^3} y \right ) \d \tau \right ]^2 + \left [  \d y + \left ( \frac{y}{r} \sqrt{\frac{r_g}{r}} - \frac{2J}{r^3} x \right ) \d \tau \right ]^2 \nonumber \\ & & + \left ( \d z + \frac{z}{r} \sqrt{\frac{r_g}{r}} \right )^2 + O(J^2)
\end{eqnarray}

\noindent ($\tau$ and $\varphi$ differ from the Schwarzschild time $t$ and polar angle $\phi$ by certain functions of $r$). This corresponds to
\begin{equation}                                                           
\vf = \nabla \chi + 2 \bJ \times \nabla \chi_1
\end{equation}

\noindent for the case $\bJ = (0, 0, J)$, where $\chi_1 = 1/r$ in the continuum; more generally, to obey (\ref{3dep-eqs}), $\chi_1$ should obey $\nabla^2 \chi_1 = 0$ at $\bx \neq {\bf 0}$. Substituting the metric into the right-hand side of (\ref{3R-g3R/2}), we find
\begin{equation}                                                           
^3 \! R^{k l} - \frac{1}{2} g^{k l} \, {}^3 \! R = 18 \frac{[\bJ \times \bn]^k [\bJ \times \bn]^l - [\bJ \times \bn]^2 n^k n^l}{r^6}
\end{equation}

\noindent as assumed equations for $g_{k l}$. As already mentioned, the iterative procedure for determining $g_{k l}$ should be performed more subtly, taking into account the dependence of the right-hand side on $g_{k l} \neq \delta_{k l}$, fulfilling three consistency conditions (that express the equality to zero of the divergence of the right-hand side). Although, it is clear that the correction to $g_{k l}$ is of the second order in $J$. The equations (\ref{6eqs}) are fulfilled at $N = 1$, $g_{k l} = \delta_{k l}$ neglecting terms of order $J^2$ (and decaying rather rapidly at $r \to \infty$). Also the left-hand side of (\ref{1eq0}) acquires a correction $O( J^2 )$ from the $\bJ$-term in $\vf$, and equation (\ref{1eq0}) remains unchanged in the linear in $J$ order. Thus, restricting ourselves to the order linear in $J$, we have similarly in the discrete case,
\begin{equation}                                                           
f_k = a^{-1} \Delta_k \chi + 2 a^{-1} \epsilon_{klm} J_l \Delta_m \chi_1 , ~~~ \sum^3_{k=1} \bDelta_k \Delta_k \chi_1 = \left \{ \begin{array}{rl} 0 & \mbox{at } \bx \neq 0  \\ 4 \pi a^2 & \mbox{at } \bx = 0 , \end{array} \right.
\end{equation}

\noindent where for $\chi$ we have
\begin{equation}                                                           
a^{-2} \sum^3_{k = 1} (\Delta_k \chi)^2 = \vf^2 (\bx ) ,
\end{equation}

\noindent where $\vf^2 (\bx )$ is given by (\ref{f^2}) (at large distances $\chi ( \bx ) = 2 \sqrt{r_g r}$), and for $\chi_1$ we have just (\ref{f^2}) (divided by $r_g$). In the formal case $r_g \lsimeqscr a$ (and $J \ll r^2_g / 4$, as in the continuum case), this is expected to give a semi-quantitative estimate for $r \lsimeqscr a$. The lattice orientation averaging applied to $\vf^2$ in equation (\ref{<f^2>}) in the non-rotating case can now also be applied to any quantity that is a function of $\vf$. At large distances, where the solution is close to the continuum one, such averaging changes this quantity little.

\section{Discussion}

Using the Schwarzschild problem as an example, we can trace the general approach to constructing discrete versions of the existing classical solutions of GR. The Einstein equations should be written in a fairly general form, not necessarily taking into account a priori symmetries that are typical for a given solution, but admitting simplicial minisuperspace geometry. The criterion for the correct choice of the required discrete solution is its approaching the continual solution at large distances. Using the periodic simplicial manifold with the 4-cubic cell divided by diagonals into 24 4-simplices and in the leading order over metric variations between the 4-simplices/4-cubes, we consider a finite-difference form of these Einstein equations.  Quantum effects show up in the elementary length scale (lattice spacing) in the zero order.

Important ingredient of the simplicial approach is averaging over various simplicial structures. Though, loose fixation of the elementary length scale $a$ already gives essential properties like resolution of the continuum GR singularities.

The length scale $a$ is defined by a free parameter $\eta$ which characterizes volume factors in the functional measure analogous to $(\sqrt{-g})^\eta$ in the continuum GR. Our analysis suggests $a \gg 1$ in Plank units.

If we take for comparison the Loop Quantum Gravity quantization of a black hole (in Kruscal coordinates) \cite{Ash1,Ash2}, an interesting feature of this quantization is that the strength of the resolved singularity (in terms of curvature) is independent of the mass of the black hole. This strength is determined by the quantum of the area spectrum, which, in turn, is determined by the Immirzi parameter $\gamma$. Note that our measure has a maximum (\ref{f^2/(d^2s/dldl)=max}) not only at the value of the elementary length scale squared $a^2$ of order of the parameter $\eta$, but also at this value of order $\gamma$. However, the maximum of the measure at the lengths squared $O(\gamma)$ is negligible compared to the maximum at the lengths squared $O(\eta)$ \cite{our1}. (It is supposed that $\eta \gg 1$ as mentioned above.) Therefore, our cut off parameter $a$ is defined by $\eta$, not by $\gamma$.

The above consideration of the Lense-Thirring metric gives hope to formulate and analyse a discrete version of the solution describing a black hole, not necessarily slowly rotating, that is, a discrete version of the full Kerr metric. A complication is, as mentioned in Subsection \ref{Lense-Thirring}, that probably we can not formulate the original continuum problem at $J \neq 0$ in terms of only $\vf$ (that is, to achieve equalities $N = 1$, $g_{k l} = \delta_{k l}$). Another case is a charged black hole and a discrete version of the Reissner-Nordström geometry, the construction of which requires incorporating the electromagnetic field in the discrete formalism. The electromagnetic field in the Regge calculus was considered in the literature \cite{Sor,Wein}.

In a more general context, we have considered finding an optimal starting point for the perturbative expansion of the theory; further, there are graviton diagrams describing quantum fluctuations around this point. In the continuum (non-renormalizable) theory, such diagrams are divergent; in the discrete framework, these diagrams are finite, and the main problem carries over to the sum of the perturbation series as in the ordinary (renormalizable) field theory. Roughly speaking, the diagrams, originally (in the continuum theory) divergent as a power of a momentum cut off $\Lambda$, are now finite and proportional to the same power of $a^{-1}$, and we have an expansion in powers of $a^{-2}$, that is, in powers of $\eta^{-1}$ at $\eta \gg 1$. An analysis of at least few first orders of this series might be of interest.

\section*{Acknowledgments}

I would like to thank R. N. Lee, A. I. Milstein, A. A. Pomeransky and participants of the Theoretical Seminar at BINP for valuable discussion. The present work was supported by the Ministry of Education and Science of the Russian Federation.

\end{document}